\documentclass[prl,aps,twocolumn]{revtex4}
\usepackage{epsfig}
\usepackage{bm}
\newcommand{\B}[1]{{\bm{#1}}}
\newcommand{\C}[1]{{\mathcal{#1}}}
\usepackage[latin1]{inputenc}
\newcommand{\beq}{\begin{equation}}
\newcommand{\eeq}{\end{equation}}
\newcommand{\bea}{\begin{eqnarray}}
\newcommand{\eea}{\end{eqnarray}}

\begin{document}
\title{Dynamic Crack Tip Equation of Motion: High-speed Oscillatory Instability}
\author{Eran Bouchbinder}
\affiliation{Racah Institute of Physics, Hebrew University of Jerusalem, Jerusalem 91904, Israel}
\begin{abstract}
A dynamic crack tip equation of motion is proposed based on the autonomy of the near-tip nonlinear zone of scale $\ell_{nl}$, symmetry principles, causality and scaling arguments. Causality implies that the asymptotic linear-elastic fields at time $t$ are determined by the crack path at a {\bf retarded time} $t\!-\!\tau_d$, where the delay time $\tau_d$ scales with the ratio of $\ell_{nl}$ and the typical wave speed $c_{nl}$ within the nonlinear zone. The resulting equation is shown to agree with known results in the quasi-static regime. As a first application in the fully dynamic regime, an approximate analysis predicts a high-speed oscillatory instability whose characteristic scale is determined by $\ell_{nl}$. This prediction is corroborated by experimental results, demonstrating the emergence of crack tip inertia-like effects.
\end{abstract}

\maketitle

{\em Introduction.}$-\!\!\!-$ Fundamental puzzles in the dynamic fracture of brittle materials remain unresolved mainly due to the lack of a well-established equation of motion for a crack's tip. The study of dynamic fracture has focused on the central idea of linear-elastic energy flowing into the crack tip nonlinear and dissipative zone \cite{98Fre,99FM}. While this approach is successful in determining the crack growth rate when its path is known a priori, it is fundamentally deficient in the general and most interesting case in which the crack's path is selected dynamically (e.g. instabilities \cite{99FM, 07LBDF}), without being supplemented with a path selection rule. This long-standing problem hampers the development of a predictive and complete theory of dynamic fracture.

In this Letter we propose a dynamic crack tip equation of motion for isotropic materials under plane deformation, based on rather general physical considerations. A basic starting point is the concept of the autonomy of the crack tip nonlinear zone in the canonical theory of fracture, linear-elastic fracture mechanics (LEFM) \cite{98Fre}. The idea is that the mechanical state within the small near-tip nonlinear zone of scale $\ell_{nl}$ (``inner problem''), where LEFM breaks down, is uniquely determined by the asymptotic linear-elastic fields surrounding it (``outer problem''), but is otherwise independent of the applied loadings and the geometric configuration (e.g. crack path) in a given problem. Therefore, the near-tip nonlinear zone is coupled to the applied loadings and the geometric configuration through the asymptotic linear-elastic fields and the crack tip itself evolves according to the dynamics within the near-tip nonlinear zone. What role then plays the near-tip nonlinear zone in determining the path selected by a crack tip?

The main idea of this Letter is that in the presence of a {\bf finite} nonlinear near-tip zone, causality implies that the asymptotic linear-elastic fields at a given time $t$, which control the crack tip motion at that time, are determined by the crack path at a {\bf retarded time} $t\!-\!\tau_d$, where the delay time $\tau_d$ scales with the ratio of $\ell_{nl}$ and the typical wave speed $c_{nl}$ within the nonlinear zone. That is, we propose that the only essential properties of the near-tip nonlinear zone are its size $\ell_{nl}$ and its typical wave speed $c_{nl}$, and that these appear in a macroscopic continuum theory mainly through their causal effect. This physical effect is missing in LEFM since its basic tenet is that the nonlinear near-tip zone acts as an energy sink, but otherwise $\ell_{nl}\!\to\!0$ can be assumed.

The mathematical formulation of these ideas leads to a simple, continuum level, dynamic equation of motion for a crack's tip. This equation agrees with the well-known ``principle of local symmetry'' \cite{74GS} in the quasi-static limit, but is shown to have novel implications in the fully dynamic regime. As an example, we perform an approximate linear stability analysis of rapid  mode I cracks. It predicts the existence of a spontaneous symmetry breaking, high-speed oscillatory instability whose characteristic scale is determined by $\ell_{nl}$. Using the recently developed weakly nonlinear dynamic fracture theory \cite{08LBF,08BLF,09BLF} to estimate $\ell_{nl}$, this prediction is shown to agree well with recent experiments \cite{07LBDF}. These results explicitly demonstrate the importance of the lengthscale $\ell_{nl}$ at high propagation speeds, as well as the emergence of crack tip inertia-like effects.

\begin{figure}
\centering \epsfig{width=.4\textwidth,file=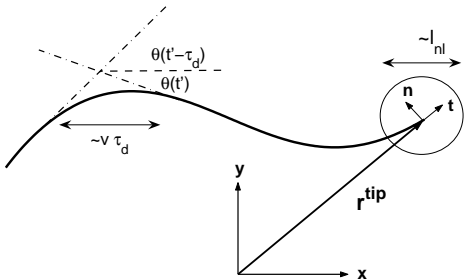} \caption{A crack with a small nonlinear zone of scale $\ell_{nl}$. The angle $\theta$ the crack makes with respect to the x-direction is shown at two times separated by a delay $\tau_d$.}\label{sketch}
\end{figure}

{\em Crack tip dynamics.}$-\!\!\!-$
The mathematical formulation of the ideas described above follows in three steps. Consider a crack under plane deformation conditions, whose path is described by $\B r^{tip}(t)$ and whose tip is surrounded by a small nonlinear zone $\ell_{nl}$, see Fig. \ref{sketch}. Note that $\ell_{nl}$ is a {\em dynamic} quantity that depends, for example, on the crack speed, $v\equiv|\dot{\B r}^{tip}|$ (cf. Figs. 1 in \cite{08LBF} and \cite{08BLF}). Outside the nonlinear zone, in an annulus whose width is a few times $\ell_{nl}(v)$, the stress tensor $\B \sigma$ is properly described by the linear-elastic universal, asymptotic fields \cite{98Fre}
\begin{equation}
\sigma_{ij}(r,\varphi,t) \simeq \frac{K_I(t) \Sigma^I_{ij}(\varphi,v) }{\sqrt{2\pi r}}+ \frac{K_{II}(t) \Sigma^{II}_{ij}(\varphi,v) }{\sqrt{2\pi r}}\ .\label{K+A}
\end{equation}
Here $r\!=\![(x-r_x^{tip})^2+(y-r_y^{tip})^2]^{1/2}$ and $\varphi\!=\!\tan^{-1}[(y-r_y^{tip})/(x-r_x^{tip})]$, where $(x,y)$ is a fixed Cartesian coordinates systems, and $i,j$ run over the polar coordinates $r$ and $\varphi$.
$K_{I,II}$ are the mode I and II stress intensity factors respectively, and $\B \Sigma^{I,II}(\varphi,v)$ are known tensorial functions \cite{98Fre}. The concept of autonomy implies that the near-tip nonlinear zone is coupled to the large scales only through the stress intensity factors $K_{I,II}$, which drive the crack's tip growth. Therefore, our aim is to derive an equation of motion based on these quantities.

As a first step in this derivation, we follow closely the reasoning of \cite{93HS}. Denote by $\hat{\bf t}$ and $\hat{\bf n}$ the tangent and normal unit vectors at the crack tip, respectively (see Fig. \ref{sketch}) and consider the discrete symmetry operation $R_n$ that transforms $\hat{\bf n}\! \rightarrow\! -\hat{\bf n}$. Under this
symmetry operation the relevant quantities of the asymptotic LEFM fields of Eq. (\ref{K+A}) transform as follows:
(i) $K_I \!\rightarrow\! K_I$ (ii) $K_{II}\! \rightarrow \!-K_{II}$ (iii) $v\! \rightarrow \!v$.
Assuming material isotropy, one can write down the
most general first order equations that are invariant under $R_n$.
The first equation is just a kinematic relation for the rate of
crack tip growth
\begin{equation}
\label{translation}
\partial_t {\bf r}^{tip} =  v(K_I(t),K_{II}(t))~\!\hat{\bf t} \ .
\end{equation}
The second one describes crack tip rotation
\begin{equation}
\label{rotation}
\partial_t \hat{\bf t} \propto K_{II}(t) \hat{\bf n} \ ,
\end{equation}
where the proportionality coefficient is a true scalar.

The second step in the derivation amounts to estimating the proportionality coefficient in Eq. (\ref{rotation}) by dimensional considerations. The existence of a lengthscale $\ell_{nl}$, the crack speed $v$ and a typical propagation stress intensity factor $\bar{K}_c$, imply that
\begin{equation}
\label{rotation1}
\partial_t \hat{\bf t} \simeq - \frac{v}{\ell_{nl}} \frac{K_{II}(t)}{\bar{K}_c} \hat{\bf n} \ .
\end{equation}
$\bar{K}_c$ can be related to the fracture energy $\Gamma(v)$, $v(K_I,K_{II})$ is determined by standard energy balance considerations, i.e. the generalized Griffith criterion \cite{98Fre}, and the minus sign ensures that cracks rotate in the proper direction in the presence of mode II fields. Our scaling approach assumes that all other material-specific properties of the nonlinear zone appear as a pre-factor of order unity in Eq. (\ref{rotation1}).
Equation (\ref{rotation1}) can be rewritten in terms of the angle $\theta$ that the unit tangent $\hat \B t$ makes with the
x-axis as \cite{03BHP, 07BP}
\begin{equation}
\label{rotation2} \partial_t \theta(t)\simeq -\frac{v}{\ell_{nl}} \frac{K_{II}(t)}{\bar{K}_c} \ .
\end{equation}

The third step in the derivation follows from the observation that $\theta(t)$ is defined at the crack tip, while $K_{II}(t)$ is defined a distance $\ell_{nl}$ away from it. Therefore, causality implies that $K_{II}$ at time $t$ cannot be affected by the crack faces created in the time interval $[t\!-\!\tau_d,t]$, with
\begin{equation}
\label{tau}
\tau_d \sim \ell_{nl}/c_{nl} \ .
\end{equation}
Here $c_{nl}$ is the typical wave speed within the nonlinear zone, possibly of the order of the linear elastic wave speed $c_s$, but not necessarily so. We note that using a single delay time $\tau_d$ is certainly a simplification of more complicated dynamics, but this simplified scaling assumption is expected to capture the essence of the physics involved.

To formulate this idea precisely, we should express the {\em physical} $K_{II}$ at time $t$ in terms of the {\em mathematical} $\C K_{II}$ at a {\bf retarded time} $t\!-\!\tau_d$, taking into account the fact that the latter is defined with respect to a coordinates system rotated by $\theta(t\!-\!\tau_d)$, while the former with respect to a coordinates system rotated by $\theta(t)$, see Fig. \ref{sketch}. $\C K_{II}(t\!-\!\tau_d)$ is obtained from a pure LEFM problem with a crack path corresponding to $t\!-\!\tau_d$. Therefore, we have
\begin{eqnarray}
\label{rotate}
K_{II}(t) \!&\simeq&\! \C K_I(t\!-\!\tau_d) \Sigma^I_{r\varphi}(\varphi=\theta(t)\!-\!\theta(t\!-\!\tau_d),v)\nonumber\\
&+&\!\C K_{II}(t\!-\!\tau_d)\Sigma^{II}_{r\varphi}(\varphi=\theta(t)\!-\!\theta(t\!-\!\tau_d),v),
\end{eqnarray}
where $\Sigma^{I,II}_{ij}(\varphi,v)$ were defined in Eq. (\ref{K+A}). The effect of the translation of the tip during the time interval $\tau_d$ may be non-negligible in general, but for the present purposes we neglect it. By combining Eqs. (\ref{rotation2})-(\ref{rotate}), we obtain the proposed dynamic equation of motion for the crack tip.

{\em The quasi-static limit.}$-\!\!\!-$ It is clear that in the quasi-static limit, $v \!\ll\! c_{nl}$, the effect of the delay time $\tau_d$ is negligible and Eq. (\ref{rotate}) predicts $K_{II}\!=\!\C K_{II}$ (use $\theta(t)\!=\theta(t\!-\!\tau_d)$, $\Sigma^I_{r\varphi}(0,v)\!=\!0$ and $\Sigma^{II}_{r\varphi}(0,v)\!=\!1$ \cite{98Fre}). Moreover, in this limit the crack tip has enough time to accommodate the presence of $K_{II}$ and thus (excluding crack initiation under imposed finite mode-mixity), we expect a predominantly mode I propagation with $\bar{K}_c\! \simeq\! K_I$. Scaling out $v$ by introducing the arc-length parametrization $s$ of the crack path, $ds\!=\!vdt$, Eq. (\ref{rotation2}) becomes
\begin{equation}
\label{QS}
\partial_s \theta \simeq - \frac{1}{\ell_{nl}}\frac{K_{II}}{K_I}  \ .
\end{equation}

The left hand side of Eq. (\ref{QS}) has the dimensions of an inverse length. In the quasi-static limit, where the effect of the delay time $\tau_d$ is negligible, this lengthscale cannot be determined by $\ell_{nl}$. Therefore, it must be determined by a macroscopic lengthscale $L$ that characterizes the sample geometry, i.e. $|\partial_s \theta|\! \sim \!L^{-1}$. The latter result is corroborated by various experimental observations, see for example \cite{85Sumi}, and is consistent with Eq. (\ref{QS}) if $K_{II}/K_I$ is of the order of $\ell_{nl}/L$. Therefore, we have
\begin{equation}
\label{PLS1}
\!\!\!|\partial_s \theta|\! \sim \!L^{-1}\Longrightarrow  K_{II}/K_I\!\sim\! \C O(\ell_{nl}/L)\! \ll \!1 \Longrightarrow K_{II}\! \approx\! 0 \ .
\end{equation}
The last result is the celebrated ``principle of local symmetry'' \cite{74GS}, which is inevitable from the LEFM perspective in which $\ell_{nl}$ does not exist. This principle, coupled to the Griffith criterion \cite{98Fre}, provides an excellent {\em quantitative} description of quasi-static crack propagation in brittle materials \cite{95ABP, 03BHP, 08PBBBJ}. In fact, in \cite{08PBBBJ} it was shown that Eqs. (\ref{QS}) and the ``principle of local symmetry'' generate indistinguishable predictions.

{\em High-speed oscillatory instability.}$-\!\!\!-$ In situations in which $v$ is of the order of $c_{nl}$, the delay time $\tau_d$ may be important, possibly giving rise to novel physical effects. To explore this possibility we use Eq. (\ref{rotation2}), with Eqs. (\ref{tau})-(\ref{rotate}), to study the linear stability of rapid mode I cracks propagating nearly steadily in a large body of size $\sim\!\!L^2$ with negligible wave interactions with the boundaries. These conditions can be easily met when an initial seed crack accelerates quickly to the center of the body \cite{07LBDF}. Thus, the crack length $l$ is $\C O(L)$.

Consider then a configuration $\C C_{\epsilon}$ that results from a
small time-dependent perturbation of the crack path $\C C_{\rm \epsilon}\!=\!\{(x,y)\!\!: -L/2 \!<\! x\! <\! vt,~ y\!=\!\epsilon \psi(x)\} \label{perturbed}$.
Here $\psi(x)$ is a smooth dimensionless function that defines transverse perturbations, $x$ is the propagation direction and $y$ is the loading direction in which a uniform tensile stress $\sigma^\infty$ is applied. The {\em dimensional} amplitude $\epsilon$ is much smaller than any other lengthscale in the problem, ensuring that both the speed and the path are only slightly perturbed. Note that we choose $l\!=\!L/2$ at $t\!=\!0$. By symmetry, $\C K_I$ and $\C K_{II}$ in the perturbed configuration have the forms
$\C K_I\! \simeq\! \C K^{(0)}_I \!+\! \C O(\epsilon^2)$ and $\C K_{II}\! \simeq \!\C K^{(1)}_{II}\! + \!\C O(\epsilon^3)$,
where the superscripts denote orders in $\epsilon$. The crack speed $v$ is quadratic in the stress intensity factors, implying that the mode II contribution to $v$ is negligible to $\C O(\epsilon)$.

We are interested in situations in which the typical scale $\lambda$ of transverse perturbations $\psi(x)$ (soon to be identified with a wavelength) satisfies $\epsilon\! \ll \!\lambda\! \ll \!l,L$. The stress intensity factors in the mathematical LEFM problem for the crack configuration $\C C_\epsilon$ were calculated in the general three-dimensional case in \cite{97WM} and applied to a two-dimensional situation in \cite{02OMW}. The result which is relevant here is \cite{02OMW, supplementary}
\begin{eqnarray}
\C K^{(1)}_{II}(t\!-\!\tau_d)&\simeq&\C K^{(0)}_I\Big(\Xi(v)-v\Theta(v)q_{II}(v)\Big)\theta(t\!-\!\tau_d)\nonumber\\
&+& \epsilon B(\psi) \ . \label{out-of-plane}
\end{eqnarray}
In this expression, as compared to the one appeared in \cite{02OMW}, we omitted a term of order $\sigma^{\infty} \epsilon /\sqrt{l}$ that is negligible with respect to $\theta \C K^{(0)}_I \!\sim\! \sigma^{\infty} \epsilon \sqrt{l}/\lambda$ ($\epsilon\! \ll \!\lambda\! \ll \!l$), where $\tan(\theta)\!=\!\epsilon\partial_x\psi(vt)\!\simeq \!\theta$.

The first term on the right-hand-side (RHS) of Eq. (\ref{out-of-plane}) is local, i.e. it depends on the angle $\theta$, and the velocity-dependent functions that appear in it are
\begin{eqnarray}
\Xi(v)\! &=&\! \partial_\varphi \Sigma^I_{r \varphi}(0,v)\! = \! \frac{8\alpha_d \alpha_s \!-\! (1
\!+\!\alpha_s^2)(\alpha_d^2\!+\!\alpha_d\alpha_s\!+\!2)}{R(v)}, \nonumber\\
\Theta(v) \!&=& \!\left[2\alpha_d (\alpha_s\!-\!\alpha_d)(1\!+\!\alpha_s^2)\right]/R(v),
\end{eqnarray}
and $q_{II}(v)$ is given explicitly in \cite{02OMW, supplementary}. $R(v)$ is the Rayleigh function \cite{98Fre,02OMW, supplementary}, $\alpha^2_{s,d}\!=\!  1\!-\!v^2/c_{s,d}^2$ and $c_s$ and $c_d$ are the shear and dilatational wave speeds, respectively. $B(\psi)$ on the RHS of Eq. (\ref{out-of-plane}) is a non-local linear functional (involving a spatio-temporal convolution). It contains a contribution that is negligible compared to the local term and a contribution that is strictly divergent \cite{02OMW, supplementary}. The latter was proposed to be regularized in terms of generalized functions \cite{02OMW} and in that case may be of the same order of magnitude as the local term \cite{supplementary}. We do not, however, consider the possibly relevant part of $B(\psi)$ here and neglect it in Eq. (\ref{out-of-plane}). We note that in the quasi-static limit $B(\psi)$ has a {\em destabilizing} role in a crack path stability analysis \cite{95ABP,03BHP}, such that its omission here may in fact promote stability. Below we show that the local terms are sufficient to induce an instability.

The resulting equation is substituted in the RHS of Eq. (\ref{rotate}), which is then expanded to first order in $\theta$ (with $\C K_I\!=\!\C K^{(0)}_I$ independent of $t$) and the result substituted in the RHS of Eq. (\ref{rotation2}). This yields the following approximation
\begin{equation}
\label{dynamic1} \partial_t\theta(t) \simeq -\frac{v}{\ell_{nl}(v)}\left[\Xi(v)\theta(t)-v\Theta(v)q_{II}(v)\theta(t-\tau_d)\right] \ ,
\end{equation}
where $\bar{K}_c\!\simeq\!\C K^{(0)}_I$ was used.

\begin{figure}
\centering \epsfig{width=.47\textwidth,file=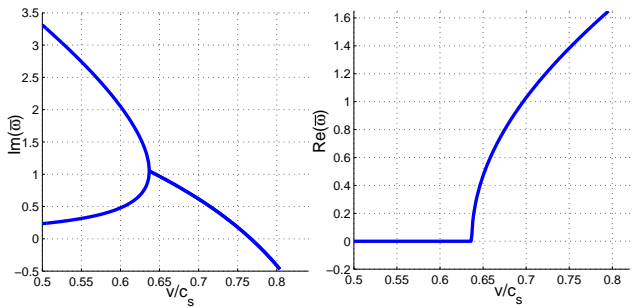} \caption{$\Im(\bar{\omega})$ (left) and $\Re(\bar{\omega})$ (right) as a function of $v/c_s$ for $\beta\!=\!0.5$, $c_{nl}\!=\!c_s$ and $c_d\!=\!2c_s$ \cite{supplementary}. A linear oscillatory instability is predicted at $v_c\!\simeq\!0.77c_s$, for which $\Im(\bar{\omega})$ becomes negative with $\Re(\bar{\omega})\!\ne\!0$. A multivalued function corresponds to multiple solutions of Eq. (\ref{complex_trans}). The symmetric $\Re(\bar{\omega})\!<\!0$ branch is not shown.} \label{instability}
\end{figure}

In order to study the linear stability of Eq. (\ref{dynamic1}), consider linear modes
$\theta(t)\!\!=\! \!a e^{i \omega t}$,
with $a\!\!\sim\!\!\epsilon/\lambda$ and $\Re{(\omega)}\!=\! 2\pi v/\lambda$. Here $\lambda$ explicitly denotes the wavelength of spatial perturbations. Substituting the linear modes into Eq. (\ref{dynamic1}), we obtain
\begin{equation}
\label{complex_trans}
i c_{nl} \bar{\omega} = -\beta v \left[\Xi(v)-v\Theta(v)q_{II}(v)e^{-i \bar{\omega}}\right] \ ,
\end{equation}
where $\bar{\omega}\!\equiv\! \omega \tau_d$ is a dimensionless complex frequency and $\beta\!\equiv\!\tau_d c_{nl}/ \ell_{nl}$ is $\C O(1)$. The solution $\bar{\omega}(v/c_s)$, which depends rather weakly on $\beta$, $c_{nl}/c_s$ and $c_d/c_s$, is shown in Fig. \ref{instability}.
The major result is that $\Im{(\bar{\omega})}$ becomes negative at a critical speed $v_c\!\simeq\!0.77c_s$ (for $\beta\!=\!0.5$, $c_d\!=\!2c_s$ and $c_{nl}\!=\!c_s$. The latter is explicitly derived in \cite{supplementary} for the material used in \cite{07LBDF}) with $\Re{(\bar{\omega})}\!\ne\!0$. This implies an oscillatory instability at $v_c$. At $v_c$ we have $\Re{(\bar{\omega}_c)}\!\simeq\!1.5$, which implies
\begin{equation}
\label{ini_wave}
\lambda = \frac{2 \pi v_c \beta \ell_{nl}(v_c)}{c_{nl}\Re{(\bar{\omega}_c)}} \simeq 1.6 \ell_{nl}(v_c) \ .
\end{equation}
The last estimate is obtained using the numbers used in Fig. \ref{instability}. To conclude, the dynamic crack tip equation of motion predicts an oscillatory instability at a high critical speed $v_c/c_s$ (weakly material dependent), with an initial wavelength $\lambda$ that scales with the linear size of the near-tip nonlinear zone $\ell_{nl}(v_c)$.

{\em Comparison to experiments.}$-\!\!\!-$ A high-speed oscillatory instability in a brittle material was observed experimentally in \cite{07LBDF}. The reported critical speed was $v_c\!\simeq\!0.87c_s$, which reasonably agrees with our theoretical prediction of $v_c\!\simeq\!0.77c_s$, especially in light of the various scaling and simplifying assumptions adopted above. In order to compare the theoretical prediction for the wavelength to the measured one, we need an estimate of $\ell_{nl}(v_c)$ in Eq. (\ref{ini_wave}). The recently developed weakly nonlinear dynamic fracture theory \cite{08BLF,09BLF} predicts nonlinear corrections to LEFM, cf. Eq. (\ref{K+A}), on a dynamic scale $\ell_{nl}(v) = h(v) G(v)/\mu$. $h(v)$ is a material-dependent function, $G(v)$ is the LEFM energy release rate that balances the fracture energy $\Gamma(v)$ \cite{98Fre} and $\mu$ is the shear modulus. Using the numbers reported in Fig. 1 of \cite{08BLF}, we estimate $\ell_{nl}(v_c)\!\simeq\!4$mm \cite{supplementary}, which upon substitution in Eq. (\ref{ini_wave}) yields $\lambda\!\simeq\!6.5$mm.

The saturated (nonlinear) wavelength $\lambda_{final}$, which cannot be computed within our linearized analysis, is reported in Fig. 4b of \cite{07LBDF} and shows a systematic variation with $\sigma^\infty$. The {\em initial} wavelength $\lambda_{ini}$ (i.e. at the onset of oscillations), however, shows no systematic variation with $\sigma^\infty$, attaining the value $\lambda_{ini}\!=\!7.5\pm2.3$mm \cite{private}. This experimental result is in agreement with our predicted $\lambda$, which pertains to the linearized dynamics near the onset of instability. Furthermore, Ref. \cite{07LBDF} reports that when the shear wave speed of the material used in the experiments was increased by a factor of 3, no appreciable change in the {\em reduced} critical speed $v_c/c_s$ was observed, in excellent agreement with the present predictions. Finally, we note that $\epsilon\!\sim\!10^{-1}$mm ($\epsilon$ corresponds to $A_{ini}$ in the notation of \cite{07LBDF}), $\lambda\simeq 7.5$mm and $L\!\sim\!10^2$mm in the experiments of \cite{07LBDF}, are all consistent with the required scales separation $\epsilon\!\ll\!\lambda\!\ll\!L$. To conclude, the theoretical predictions of the proposed theory are in agreement with the available experimental data.

{\em Summary.}$-\!\!\!-$ We have proposed a dynamic equation of motion for crack tips propagating in isotropic materials under plane deformation.
A crucial ingredient in the derivation is the idea that a finite near-tip nonlinear zone $\ell_{nl}$ introduces a delay time $\tau_d$ between the driving stress intensity factors and the crack tip itself. The resulting equation agrees with known quasi-static results and predicts a high-speed oscillatory instability with a characteristic scale $\ell_{nl}$, in the fully dynamic regime. The latter result shows explicitly that the non-geometric, dynamic lengthscale $\ell_{nl}(v)$ directly influences the crack's dynamics at high speeds, cf. \cite{07BP}. Furthermore, the existence of a finite $\ell_{nl}(v)$ induces crack tip inertia-like effects associated with the delay time $\tau_d$. All of these effects are missing in LEFM \cite{98Fre,99FM}.

When the present predictions are supplemented with the predictions of the recently developed weakly nonlinear fracture theory for $\ell_{nl}(v)$, a satisfactory agreement with experimental results for the onset of a high-speed oscillatory instability is obtained. Therefore, extending the present ideas to three-dimensional situations may open the way to understanding the side-branching instability, which also involves a poorly understood lengthscale \cite{99FM}.

\end{document}